\documentclass{ws-procs961x669}
\usepackage{graphicx}
\usepackage{color}

\newcommand{\citen}[1]{\citenum{#1}}
\newcommand{\Eq}[2][Eq.~]{#1(\ref{eq:#2})}
\newcommand{\Fig}[2][Fig.~]{#1\ref{fig:#2}}
\newcommand{\D}{\mathrm{d}}
\newcommand{\Ai}{\mathrm{Ai}}

\makeatletter
\renewcommand{\ps@plain}{%
  \renewcommand{\@oddhead}{\hfil{\footnotesize%
    A contribution to the Julian Schwinger Centennial Conference, %
    7--12 February 2018, Singapore}\hfil}%
  \renewcommand{\@evenhead}{\@oddhead}%
  \renewcommand{\@oddfoot}{\hfil\thepage}%
  \renewcommand{\@evenfoot}{\thepage\hfil}%
}
\makeatother
\crop[off]

\begin{document}

\title{Julian Schwinger and the Semiclassical Atom} 

\author{Berthold-Georg Englert}
\address{Centre for Quantum Technologies\\%
  and Department of Physics, National University of Singapore;\\%
  MajuLab, Singapore;\\%
  {cqtebg@nus.edu.sg}} 

\begin{abstract}
In the early 1980s, Schwinger made seminal contributions to the semiclassical
theory of atoms. 
There had, of course, been earlier attempts at improving upon the
Thomas--Fermi model of the 1920s. 
Yet, a consistent derivation of the leading and next-to-leading corrections to
the formula for the total binding energy of neutral atoms,
\begin{displaymath}
        -\frac{E}{e^2/a_0}=0.768745 Z^{7/3}-\frac{1}{2} Z^2
                          +0.269900 Z^{5/3}+\cdots\,,
\end{displaymath}
had not been accomplished before Schwinger got interested in the
matter;
here, $Z$ is the atomic number and $e^2/a_0$ is the Rydberg unit of energy.
The corresponding improvements upon the Thomas--Fermi density were next on his
agenda with, perhaps, less satisfactory results.
Schwinger's work not only triggered extensive investigations by
mathematicians, who eventually convinced themselves that Schwinger got it
right, but also laid the ground, in passing, for later refinements --- some of
them very recent. 
\end{abstract}

\section{Introduction}
Julian Schwinger's groundbreaking work on quantum electrodynamics first and
then, more generally, on quantum field theory and the physics of elementary
particles is, of course, at the center of his legacy.
In addition, he made seminal contributions to many other topics, among them
the semiclassical theory of atoms.
This work of his is summarized in an essay of 1985 for non-expert readers
that did not get published in his lifetime.\cite{1985essay}
I shall recall here some aspects of our collaboration from mid 1981 to early
1985 and also mention later work that was triggered by Schwinger but did not
involve him.

\section{Schwinger's papers of 1980 and 1981}
\begin{figure}[!t]
  \centerline{\includegraphics[viewport=130 480 400 690]%
               {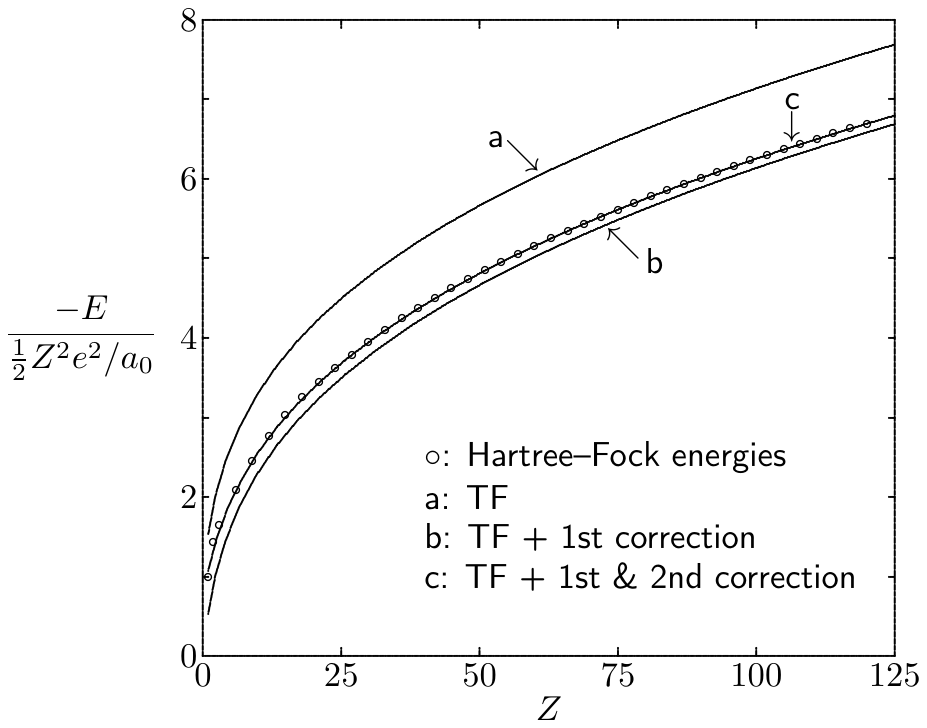}}
  \caption{\label{fig:HFvsTFSS}%
    Total binding energy of neutral atoms, as a function of the atomic number
    $Z$, in Rydberg units ($e^2/a_0$) and scaled by $\frac{1}{2}Z^2$.
    The circles represent the Hartree--Fock values, and curves \textsf{a},
    \textsf{b}, and \textsf{c} show, respectively, the Thomas--Fermi
    approximation by itself, and with the leading and the next-to-leading
    corrections included. See also Fig.~8 in Ref.~\citen{1985essay}.}  
\end{figure}

While teaching a course on quantum mechanics in the late 1970s, Schwinger was
intrigued by the Thomas--Fermi (TF) model and the systematic deviation of the
TF approximation for the total binding energy of an atom from the
Hartree--Fock (HF) values (curve \textsf{a} and circles in \Fig{HFvsTFSS}).
This was crying out for an understanding.
Schwinger responded to the challenge with his derivation (in
1980\cite{JS-1980}) of the leading correction that had been conjectured by
John Scott in 1952,\cite{Scott-1952} now known as the \emph{Scott correction}.
Schwinger did not know Scott's paper before a referee pointed it out.

Although the inclusion of the Scott correction improved the binding energies
substantially (curve \textsf{b} in \Fig{HFvsTFSS}), there remained a
systematic error.
In a paper of 1981,\cite{JS-1981} Schwinger took care of this as well, again
giving a clear-cut demonstration of what had been conjectured
before;\cite{P+D-1978} this takes us to curve \textsf{c} in \Fig{HFvsTFSS}
that goes right through the circles of the HF values.
The consecutive approximations are the three terms in Eq.~(46) of
Ref.~\citen{1985essay}, which are recalled in the Abstract.

\section{Mathematicians in action}
Schwinger's derivation of the two leading corrections to the TF energy of
atoms in Refs.~\citen{JS-1980} and \citen{JS-1981} convinces any theoretical
physicist, but the standards of mathematical physics are different.
Earlier, in 1977, Elliot Lieb and Barry Simon had given a proof that the TF
approximation is asymptotically exact as $Z\to\infty$; see
Ref.~\citen{L+S-1977} for the precise statement.
Now, two other duos of mathematicians dealt with Scott's leading correction
and Schwinger's next-to-leading correction.

First, Heinz Siedentop and Rudi Weikard in a series of
papers\cite{S+W-1986,S+W-1987a,S+W-1987b,S+W-1987c,S+W-1989}
published 1986--1989 showed that Scott's correction is indeed the leading
correction when $Z$ is large enough.
Then, Charles Fefferman and Luis Seco devoted another series of
papers\cite{F+S-1989,F+S-1990a,F+S-1990b,F+S-1992,F+S-1993,F+S-1994a,%
  F+S-1994b,F+S-1994c,F+S-1995} to the second correction, published
1989--1995, eventually confirming that Schwinger's result is correct.

These three important pieces of mathematical research illustrate how, in the
history of this subject matter, the link between theoretical physics and 
mathematics has been a one-way road: The physicists provide conjectures
and the mathematicians convert them into theorems.
There hasn't been any benefit for the physicists in return,
beyond being assured that they got it right (which, by their own standards,
they knew already).
The work by Lieb and Simon did not yield a suggestion on how to improve on the
TF approximation, nor did the Siedentop--Weikard proof indicate how to go
beyond the Scott correction, and an analogous remark applies to the
accomplishments by Fefferman and Seco.
While it is true that the task chosen by each of the three duos was limited to
proving what was conjectured, it would have been also nice to get an idea
about where to look for the next improvement.

\section{Summer of 1981: A homework assignment }
Schwinger finished the work on the corrections to the TF energy by the end of
1980 (Ref.~\citen{JS-1981} was submitted in December 1980), and then began
studying the corresponding modifications of the TF density.
Lester DeRaad had already provided numerical solutions to a modification of
the TF differential equation,\cite{JS-1981,DeR+S-1982}
\begin{equation}\label{eq:1}
  \frac{\D^2\phi}{\D y^2}=\frac{\phi^{3/2}}{y^{1/2}}+\phi
\end{equation}
with
\begin{equation}\label{eq:2}
  \phi(0)=\frac{48\pi}{(22)^{3/2}}Z=1.461 Z
  \quad\text{and}\quad\phi(\infty)=0\,,
\end{equation}
where $\phi(y)$ is the auxiliary function in terms of which the quantities of
interest can be computed; see Ref.~\citen{DeR+S-1982} for details.
As expected, this improved on the TF approximation for the electron density
at large distances --- the values obtained for diamagnetic susceptibilities,
essentially the squared distance from the nucleus weighted by the density,
were much better than the TF values --- but, so the paper summarizes, the
``outcome of this test [\dots] hardly warrants proclaiming a successful
conclusion to the search for an extrapolation of the TF model into the outer
regions of the atom.''  

\begin{figure}[t]
  \centerline{\includegraphics{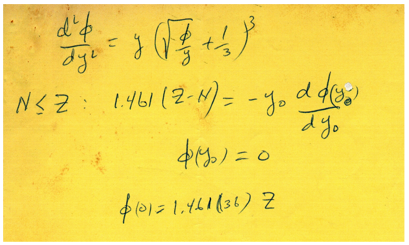}}
  \caption{\label{fig:homework}The homework assignment: Schwinger's
    modification of the TF differential equation in his hand writing.
    The white spot is a drop of correction fluid.}
\end{figure}

Accordingly, this search continued and led Schwinger to considering
\begin{equation}\label{eq:3}
  \frac{\D^2\phi}{\D y^2}
  =y{\left(\sqrt{\frac{\phi}{y}}+\frac{1}{3}\right)}^3
\end{equation}
with
\begin{equation}\label{eq:4}
  \phi(0)=1.461 Z
  \quad\text{and}\quad \phi(y_0)=0\,,\quad
    1.461(Z-N)=-y_0\frac{\D\phi(y_0)}{\D y_0}\,,
\end{equation}
which he gave to me as a homework assignment in the summer of 1981 (see
\Fig{homework}).
At this time, he spent part of a sabbatical leave in T\"ubingen,
hosted by my Ph.D. supervisor Walter Dittrich.
I had recently submitted my thesis and was waiting for the defense, and
so I was available when Schwinger needed someone with programming skills.
It was easy to get the numbers, such as a table of $y_0$ values for different
$Z$ and $N$, and the numbers passed the various tests that Schwinger could
subject them to.

He then gave me a copy of the draft paper he was working on, and I finally
understood the origin of \Eq{3} and the physical meaning of $\phi(y)$.
While my feedback on the draft and other input did have a bearing on the
published version,\cite{E+S-1982} it was hardly enough to earn co-authorship.
Schwinger, however, thought differently and my objection to being first author
were gently and firmly pushed aside: ``There is much to be said in favor of
the alphabet.''

The electron densities from the formalism based on \Eq{3} do improve on the TF
approximation.
They give reasonably good values for the diamagnetic susceptibilities,
better than the HF numbers for larger atoms,%
\footnote{In those days, HF numbers were systematically too large by about
  10\%.} 
and this was good motivation for taking the next step.

\section{1984 and 1985: Two trios of papers}
By the time Ref.~\citen{E+S-1982} was finalized and submitted, I had been
recruited by Schwinger as a postdoc, and we were already working on the
follow-up project that yielded a trio of papers in
1984.\cite{E+S-1984a,E+S-1984b,E+S-1984c}

\Eq[Equations~]{1} and \Eq[]{3} were modifications of the TF equation that
incorporated Paul Dirac's approximate exchange energy\cite{Dirac-1930} and 
Carl Friedrich von Weizs\"acker's inhomogeneity correction,\cite{vW-1935} both
responsible for the second correction, but not Scott's leading correction.
While this is acceptable when the focus is on the outer reaches of the atom,
it is somewhat inconsistent.

Scott's correction addresses the failure of the TF approximation in the
vicinity of the nucleus where the singularity of the Coulomb potential results
in large changes of the potential energy over short distances.
Therefore, it was necessary to treat the strongly bound electrons
differently.\cite{E+S-1984a}
While Dirac's exchange energy could be used without any essential modification,
Weizs\"acker's inhomogeneity correction required a refinement that
accounts for higher-order terms, and Schwinger designed an ingenious method
for that, where a certain averaging%
\footnote{Strictly speaking, it is not an average because the terms are
  weighted by the Airy function $\Ai(x)$, which is assuredly positive for
  ${x>0}$ but not for ${x<0}$.
  See Ref.~\citen{2D-Airy} for a recent benchmarking exercise.}
over simpler Weizs\"acker-type expressions is performed.\cite{E+S-1984b}
An extensive numerical study confirmed that the resulting modification of the
TF model was worth the effort.\cite{E+S-1984c}

It is, however, not truly satisfactory because of an unresolved problem with
the handling of the strongly bound electrons.
That involves a separation on the energy scale, and the electrons with a
binding energy less than the threshold value are treated in the TF fashion
(with modifications).
The electrons with larger binding energy are regarded as dominated by the
Coulomb potential of the nucleus, with corrections accounted for by
perturbation theory.
As a consequence of the rather different treatments, there is an unphysical
dependence on the choice of the threshold value.
While this can be systematically removed in the resulting correction to the
energy (this is an important detail in Schwinger's derivation of Scott's
correction\cite{JS-1980}), we did not manage to get rid of it in the density.%
\footnote{The situation is remarkably different in momentum space where the
  Scott-corrected density does not suffer from this
  problem.\cite{p-TFS1,p-TFS2}} 

Meanwhile, we were wondering about the difference between the HF energies in
\Fig{HFvsTFSS} and those of curve \textsf{c}.
As Figs.~9 and 10 in Ref.~\citen{1985essay} show, this difference is a
not-so-regular oscillatory contribution, which suggests that the degeneracy
associated with energy shells --- Niels Bohr's shells\cite{Bohr-1913} modified
by the repulsive forces between electrons --- is important here. 
This suggestion was not misleading and eventually provided an understanding of
these energy oscillations, reported in another trio of papers, published in
1985.\cite{E+S-1985a,E+S-1985b,E+S-1985c}. 

Here, a crucial observation is that the TF approximation can be obtained by
first expressing the energy as a sum over WKB energies, and then replacing the
sum by an integral.\cite{E+S-1985a}
This integral is a zeroth-order approximation of the sum, and the hierarchy of
higher-order terms contains oscillatory terms that can be extracted
systematically. 
The leading oscillatory term contributes\cite{E+S-1985c}
\begin{eqnarray}\label{eq:5}
  &&- 0.4805 \,Z^{4/3}\frac{e^2}{a_0}
  \sum^\infty_{k=1} \frac{(-1)^k}{(\pi k)^3}
  \sin (2\pi k\lambda_0)\nonumber\\
  &=&0.3206\,\lceil\lambda_0\rfloor
  {\left(\frac{1}{4}-\lceil\lambda_0\rfloor^2\right)}
  \,Z^{4/3}\frac{e^2}{a_0}
\end{eqnarray}
to the energy, where $\lambda_0=0.928Z^{1/3}$ and $\lceil \lambda_0\rfloor$
denotes the difference between $\lambda_0$ and its nearest integer;
see also Eq.~(63) in Ref.~\citen{1985essay}.
After the TF energy ($\propto Z^{7/3}$), Scott's leading correction
($\propto Z^{6/3}$), and Schwinger's second correction ($\propto Z^{5/3}$),
this contribution ($\propto Z^{4/3}$) is the next in line.
As of today (2019), there is no mathematical proof of its correctness, and so
we have to be content with the evidence presented in Ref.~\citen{E+S-1985c}.

Consult Ref.~\citen{LNP300} for a detailed account of all these developments.


\section{Functionals of both the density and the effective potential}
\begin{figure}[t]
  \centerline{\includegraphics{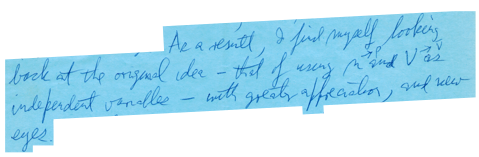}}
  \caption{\label{fig:JS-1981}Schwinger's appreciation, in 1981, of
    functionals with both the density and the potential as
    independent variables.}
\end{figure}

The argument presented in Ref.~\citen{E+S-1982} proceeds from a functional for
the energy that has the single-particle density, the electrostatic potential,
and the chemical potential as independent variables. 
Schwinger wrote this down without giving a derivation but justifying it by its
consequences, and I must have just accepted it.
It is indeed consistent with a remark by Schwinger in a letter of 1981 that he
wrote in response to the detailed feedback I gave on the draft paper mentioned
above; see \Fig{JS-1981}.
I~am pretty sure that I did not appreciate the flexibility of this approach at
the time.
In hindsight, I regard the functional in Ref.~\citen{E+S-1982} as very clever
and just right for the purpose at hand, but not as a useful starting point for
more general developments. 

We did not use anything quite like this again and, instead, switched to a
formalism based on energy functionals of the particle density, the
\emph{effective} potential, and the chemical potential as independent
variables.  
The effective potential is the sum of the external potential (here: the Coulomb
potential of the nucleus) and an interaction potential obtained from the
response of the interaction energy (here: direct and exchange electrostatic
energy) to variations of the density.
This was a natural thing to do, it was intuitive, and no justification seemed
necessary --- it was clearly an obvious matter for Schwinger, an example of
his phenomenal intuition about physics and its mathematical language.

As for myself, I took some time to comprehend fully what was so obvious to him.
Eventually, at the time of writing Ref.~\citen{p-TF}, with a precursor in
Ref.~\citen{LNP300}, I saw how this is systematically connected to the
standard density-functional theory as formalized by Pierre Hohenberg and
Walter Kohn\cite{H+K-1964} --- and this turned the density-potential
functional into a central tool for all of my subsequent work on related topics. 
The connection is actually quite simple: just subject the kinetic energy term
in the Hohenberg--Kohn density functional to a Legendre transformation.

The concept of density-potential functionals is a contribution of lasting
value, perhaps Schwinger's most important contribution to the field.
These functionals are more flexible than density-only functionals, and they
facilitate systematic approximations that go beyond formal gradient
expansions, for which the Weizs\"acker term of 1935\cite{vW-1935} is the
prototype. 
In particular, there is a density-potential functional for the Scott-corrected
TF model but no density-only functional.

As another example, I mention the somewhat amusing case of gradient
corrections to the TF model in two dimensions.
Papers in the early 1990s established --- or so it seems --- that there are no
such corrections, all terms in a formal power series of the density gradient
have vanishing coefficients; see, for example, Refs.~\citen{Holas+2:91} and
\citen{Shao:93}.
On the other hand, the TF approximation to the kinetic energy is not exact,
whether in one, two, or three dimensions, and so one has the puzzling
situation of a non-exact approximation with all systematic correction terms
vanishing. 
The puzzle disappears as soon as one recognizes that there are nonzero
gradient corrections in the density-potential functional, and they can be
evaluated perturbatively.\cite{2DvW}

There is also an analog of the density-potential functionals in momentum
space, where one gets an energy functional with the momentum-space
density, the effective kinetic energy, and the chemical potential as
independent variables.
The hierarchy is repeated:
We have the momentum-space version of the TF model\cite{p-TF} (which has
succumbed to the power of mathematics\cite{p-TF-math}), of the Scott-corrected
TF model,\cite{p-TFS2} and of a model with exchange energy and gradient
corrections included as well.\cite{p-TFSS}

Finally, there is a recent, novel attempt at approximating the potential
part of the density-potential functional.
There is a connection with the single-particle propagator that was already
exploited in Ref.~\citen{E+S-1984b} and led then to the Airy-averaged
expressions mentioned above, with an improvement over a related method by
Eugene Wigner\cite{Wigner-1932} and John Kirkwood.\cite{Kirkwood-1933}
Now, one can alternatively approximate the propagator by a factorization into
terms that refer only to the kinetic energy or only to the potential energy
--- a technique introduced and developed by Hale Freeman Trotter\cite{Trotter}
and Masuo Suzuki\cite{Suzuki} 
--- and this yields very good approximations without a gradient 
expansion.\cite{PropagatorMethod} 
The applications thereof are work in progress.

\section{Summary}
Schwinger's papers of 1980 and 1981 on the leading and the second correction to
the TF energy are well known and have triggered extensive mathematical studies.
His insight that one should treat the density and the effective potential as
independent variables on equal footing is just as important and crucial for
ongoing research.

\section*{Acknowledgments}
Julian Schwinger, always kind and generous, taught and guided me from 1981
till his premature death in 1994.
His lessons, both formal and informal, were invaluable.
I was a greenhorn when he recruited me as his postdoc at UCLA.
When I moved to the University of Munich forty months later,
I was a physicist.

\end{document}